\documentclass[a4paper,twocolumn]{article}%
\usepackage{amsfonts}
\usepackage{scicite}
\usepackage{times}
\usepackage{amsmath}
\usepackage{amssymb}
\usepackage{graphicx}%
\providecommand{\U}[1]{\protect\rule{.1in}{.1in}}
\newenvironment{sciabstract}{\begin{quote} \bf}
{\end{quote}}

\newcounter{lastnote}

\newtheorem{theorem}{Theorem}
\newtheorem{acknowledgement}[theorem]{Acknowledgement}

\begin{document}

\title{Extreme Nonlinear Phenomena in NbN Superconducting Stripline Resonators}
\author{Eran Arbel-Segev$^{\ast}$,Baleegh Abdo, Oleg Shtempluck, Eyal Buks\\\ \\Microelectronics Research Center, Department of Electrical Engineering,\\Technion, Haifa 32000, Israel\\{\normalsize {$^{\ast}$To whom correspondence should be addressed; E-mail:
segeve@tx.technion.ac.il.} }}
\date{}
\maketitle

\begin{sciabstract}
We study a microwave superconducting stripline resonator made of NbN on a
Sapphire wafer. Novel, self-sustained modulation of the reflected power off
the resonator, at frequencies of up to 60MHz, has been recently reported. Here
we show experimentally that near the self modulation threshold, the device
exhibits a giant nonlinearity, which manifests itself in an extremely high
intermodulation gain, accompanied by a very strong noise squeezing and period
doubling of various orders. Such a device is highly suitable for serving as a
readout device in quantum data processing systems.
\end{sciabstract}

The theory of nonlinear dynamics predicts that near the edge of instability,
one can expect a strong noise amplification, known also as noise rise
\cite{bifAmp_Wiesenfeld85}, which is linearly unbounded and only saturates by
higher order nonlinear terms \cite{BifAmp_Kravtsov03}. In addition, the same
mechanism can result in large amplification of small periodic signals,
injected into the system\cite{BifAmp_Wiesenfeld86}. Nonlinear effects in
superconductors have significant implications for both basic science and
technology. It may be exploited to explore some important quantum phenomena in
the microwave region, such as quantum squeezing \cite{Sqz_Movshovich90,
Squeezing_Yurke05, SQZ_Buks05} and experimental observation of the so called
dynamical Casimir effect\cite{segev06e}. Whereas technologically, these
effects may allow some intriguing applications such as bifurcation amplifiers
for quantum measurements \cite{BifAmp_Siddiqi04,BifAmp_Wiesenfeld86}, resonant
readout of qubits \cite{QBITS_Lee05}, mixers \cite{HEB_Floet99}, single photon
detectors \cite{HED_Goltsman05}, and more. Recently, we have reported on a
novel nonlinear phenomenon originates by thermal instability, in which
self-sustained modulation (SM) of a reflected pump tone off a resonator is
generated in a superconducting stripline resonator \cite{segev06b}. At the SM
power threshold range the resonator experiences strong noise
amplification\cite{segev06c}. A theoretical model, which is presented in Ref.
\cite{segev06c}, exhibits a good agreement with the experimental results.

In this paper we experimentally demonstrate the significance of the SM
phenomenon, as a generator of several giant nonlinear effects, which manifest
themselves at the instability threshold. First, a very strong amplification of
periodic signals is demonstrated using intermodulation (IM) measurement. The
same measurement technique is used to show period doubling bifurcation (PDB)
of various orders. Finally, we demonstrate a phase sensitive deamplification
(PSD) which exhibits a strong squeezing factor. Moreover, in the same region
our devices have already proved an ability to amplify the mixing products of a
low power optical signal, modulated at gigahertz frequencies, and a RF signal,
with a large signal to noise ratio\cite{Segev06a,segev06e}.
\begin{figure*}
\centering
\[
\text{%
{\includegraphics[
height=1.276in,
width=4.3653in
]%
{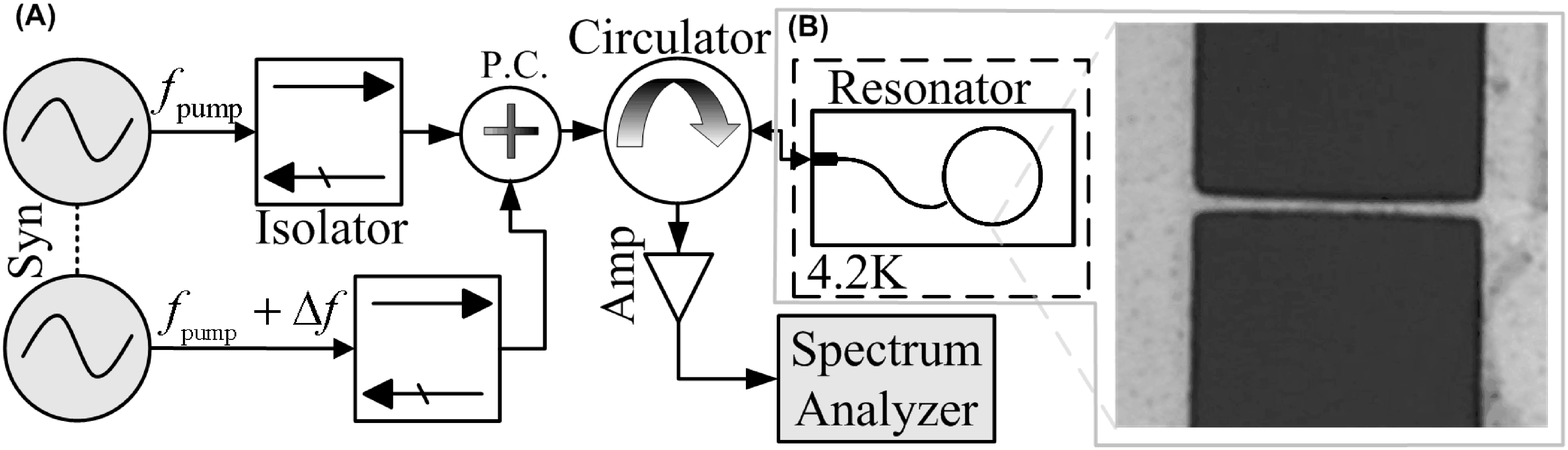}%
}%
}%
\]
%

\caption{
$(\mathrm{A}%
)$ IM measurement setup. The resonator is stimulated by two phase-locked tones. One,
called Pump, is a relatively intense tone which biases the resonator into the
nonlinear region. The other, called Signal, is a relatively weak tone,
namely at list three orders of magnitude weaker that the pump tone.
The reflected power off the resonator is measured using a spectrum
analyzer (SA). $(\mathrm{B})$ The resonator is made as a stripline ring,
having a characteristic impedance of $50\operatorname{\Omega}$. It is
composed of $200\operatorname{nm}$ thick Niobium Nitride (NbN) deposited on a
Sapphire wafer. A weakly coupled feedline is employed for delivering the input
and output signals. The first few resonance frequencies fall within the range
of $2-8\operatorname{GHz}$. The optical image shows the microbridge whose
dimensions are $1\times10\operatorname{\mu m}^{2}$.}%
\label{ExpSetup}%
\end{figure*}%

Our experiments are performed using a novel device (Fig. \ref{ExpSetup}%
$\left(  \mathrm{B}\right)  $), which integrates a narrow microbridge into a
superconducting stripline ring resonator, and consequently has resonance
frequencies, which can be tuned by both internal (Joule self heating) or
external (infrared illumination) perturbations. The microbridge serves as a
self-tuned, lumped element, which governs the boundary conditions of the
resonator \cite{supRes_Saeedkia05}, and thus allows the manipulation of its
resonance frequencies. Further design considerations, fabrication details as
well as normal modes calculation can be found elsewhere \cite{Segev06a}. The
IM experiments are performed using the setup described in Fig. \ref{ExpSetup}%
$(\mathrm{A})$ while the device is fully immersed in liquid Helium.
\begin{figure*}
\centering
\[%
\begin{array}
[c]{ccc}%
\text{%
{\parbox[b]{2.0714in}{\begin{center}
\includegraphics[
height=1.6181in,
width=2.0714in
]%
{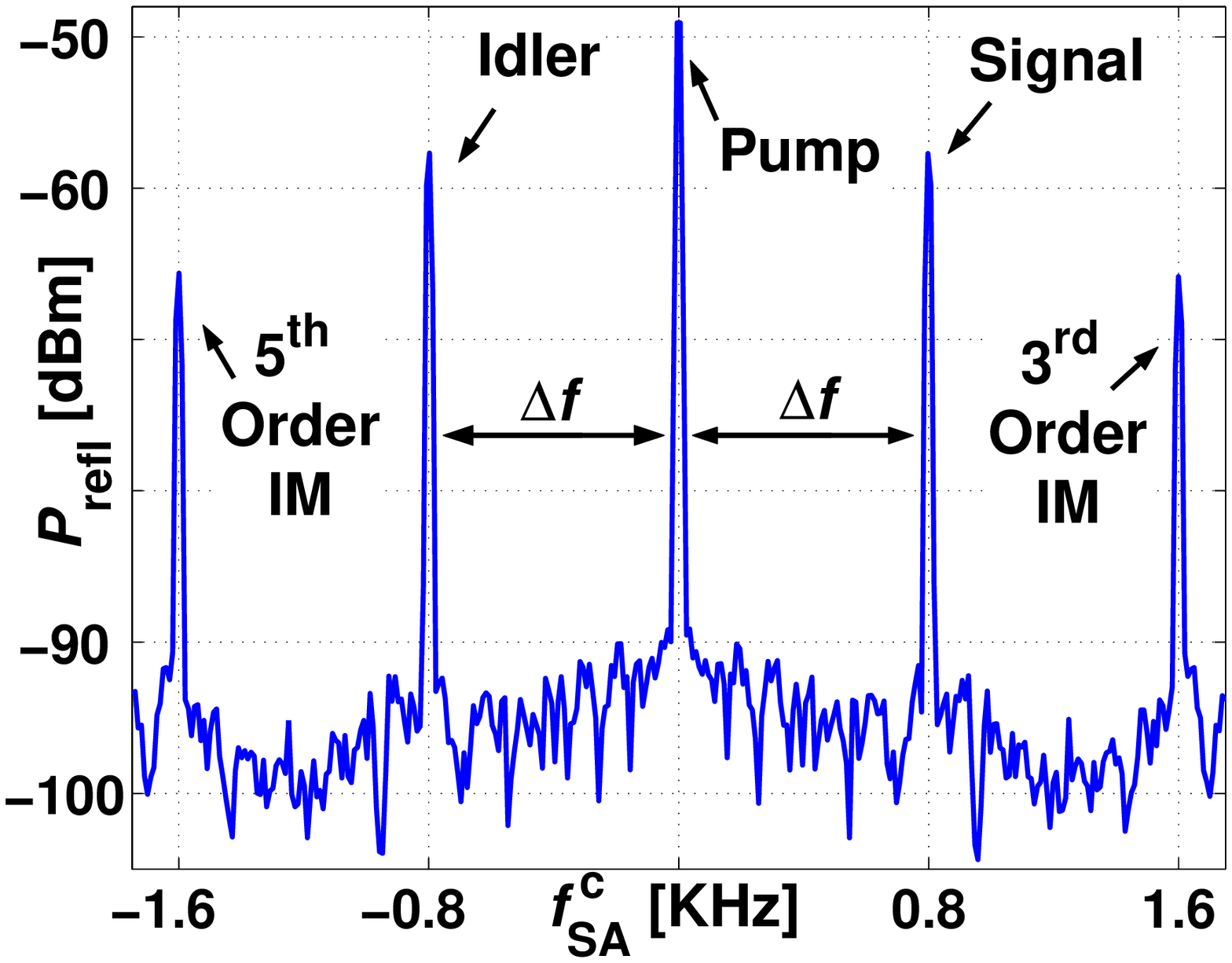}%
\\
{\protect\small (A)}%
\end{center}}}%
} & \text{%
{\parbox[b]{2.0714in}{\begin{center}
\includegraphics[
height=1.6164in,
width=2.0714in
]%
{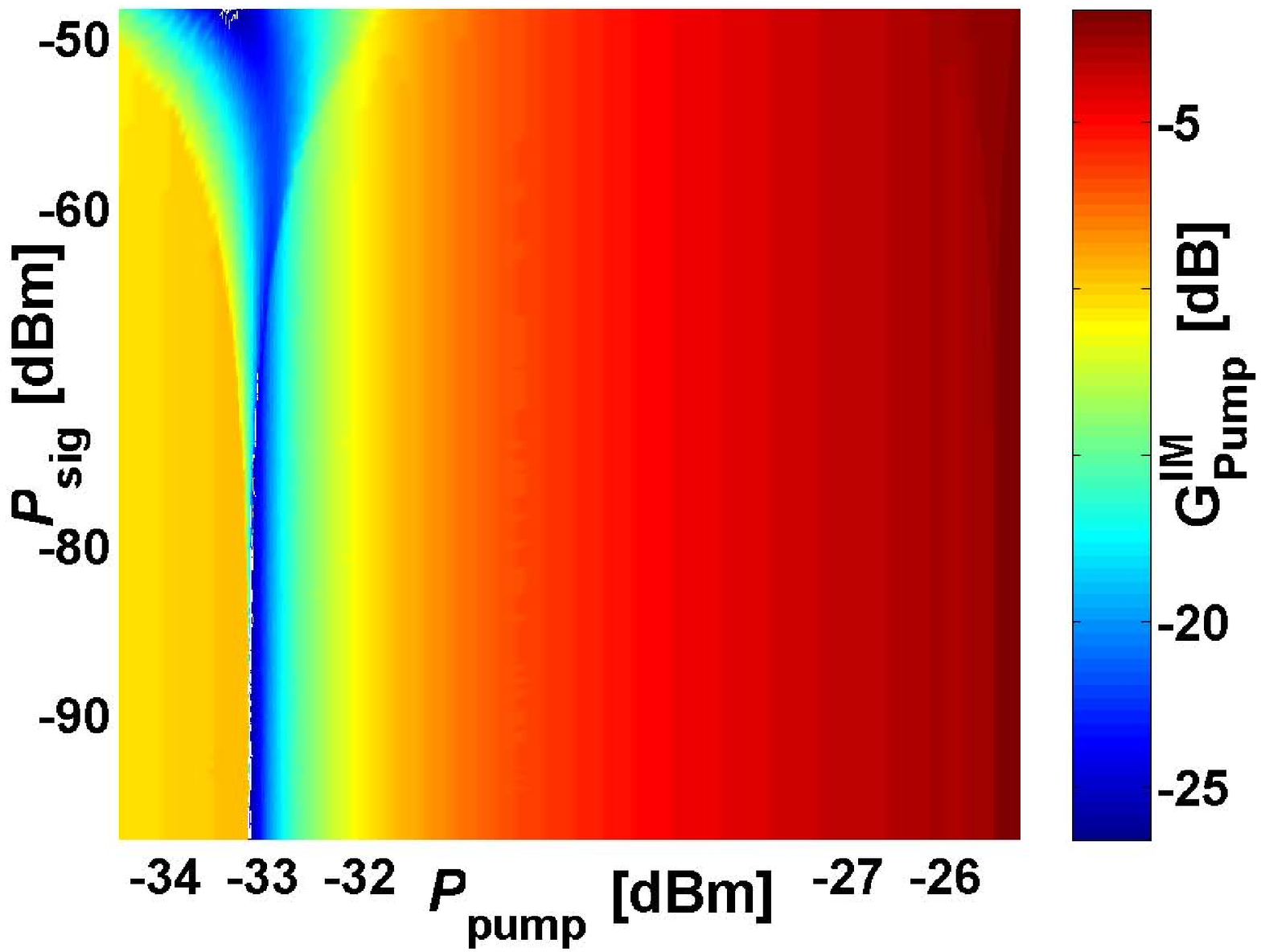}%
\\
{\protect\small (B)}%
\end{center}}}%
} & \text{%
{\parbox[b]{2.0714in}{\begin{center}
\includegraphics[
height=1.6289in,
width=2.0714in
]%
{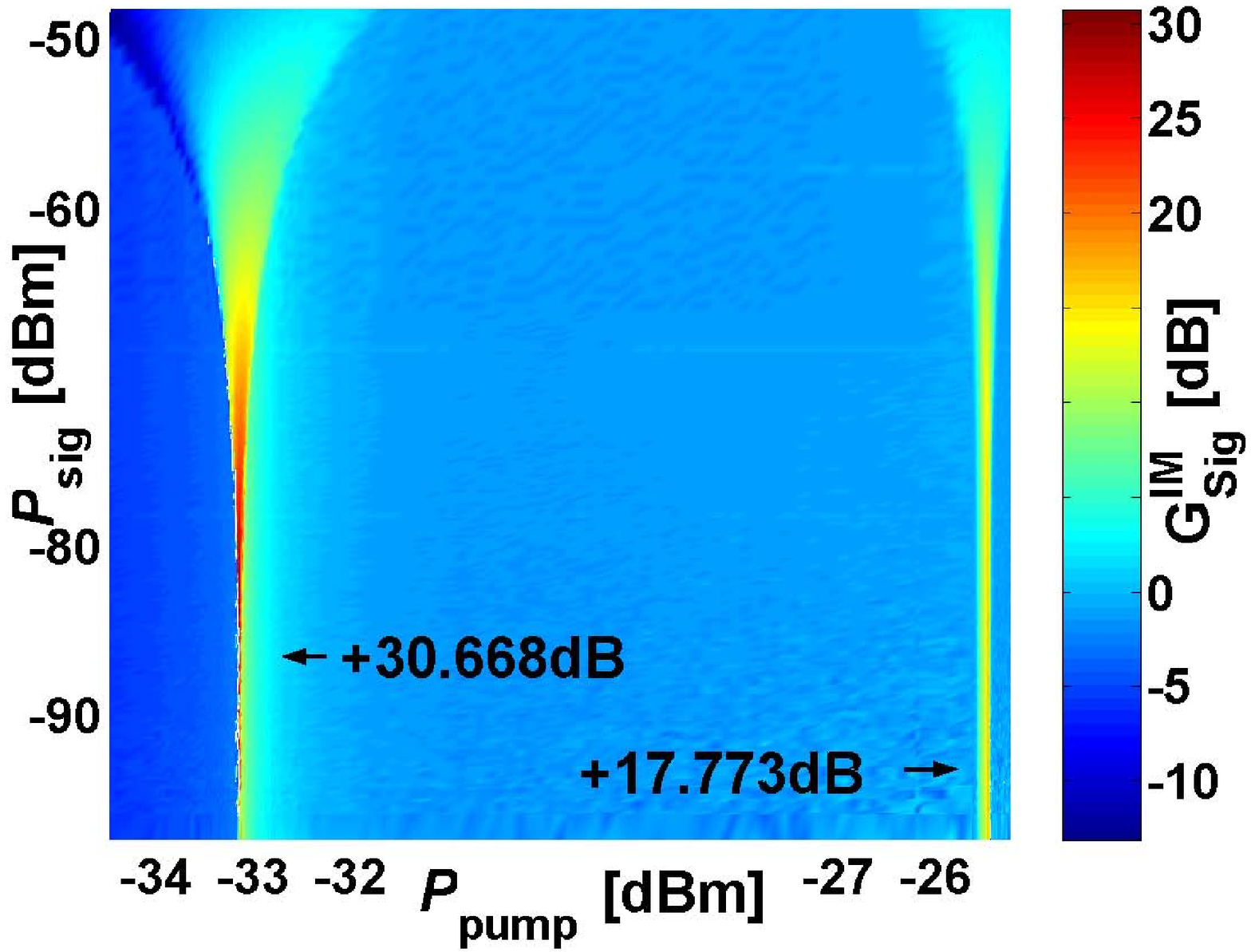}%
\\
{\protect\small (C)}%
\end{center}}}%
}\\
\text{%
{\parbox[b]{2.0714in}{\begin{center}
\includegraphics[
height=1.6272in,
width=2.0714in
]%
{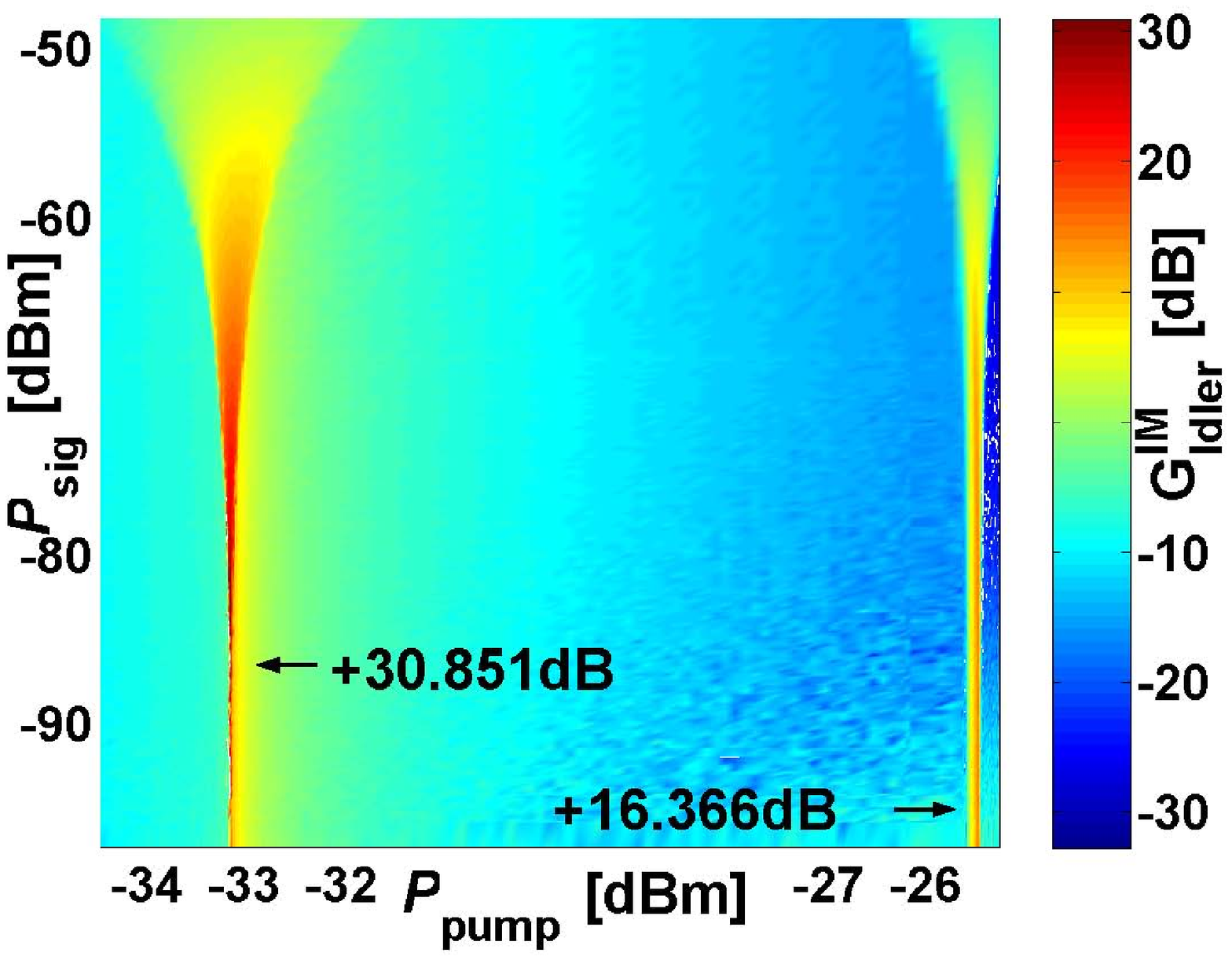}%
\\
{\protect\small (D)}%
\end{center}}}%
} & \text{%
{\parbox[b]{2.0714in}{\begin{center}
\includegraphics[
height=1.6189in,
width=2.0714in
]%
{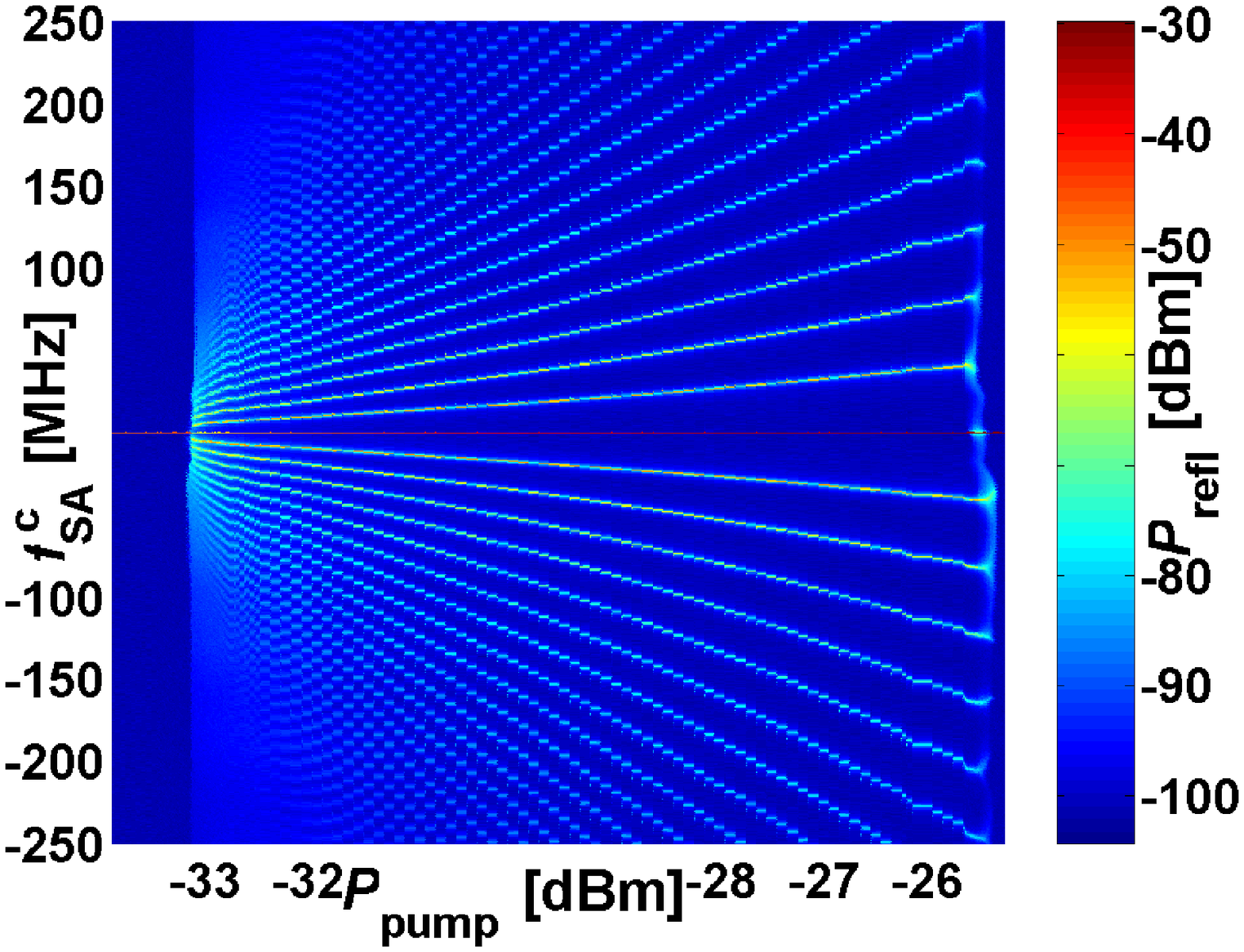}%
\\
{\protect\small (E)}%
\end{center}}}%
} & \text{%
{\parbox[b]{2.0714in}{\begin{center}
\includegraphics[
height=1.6181in,
width=2.0714in
]%
{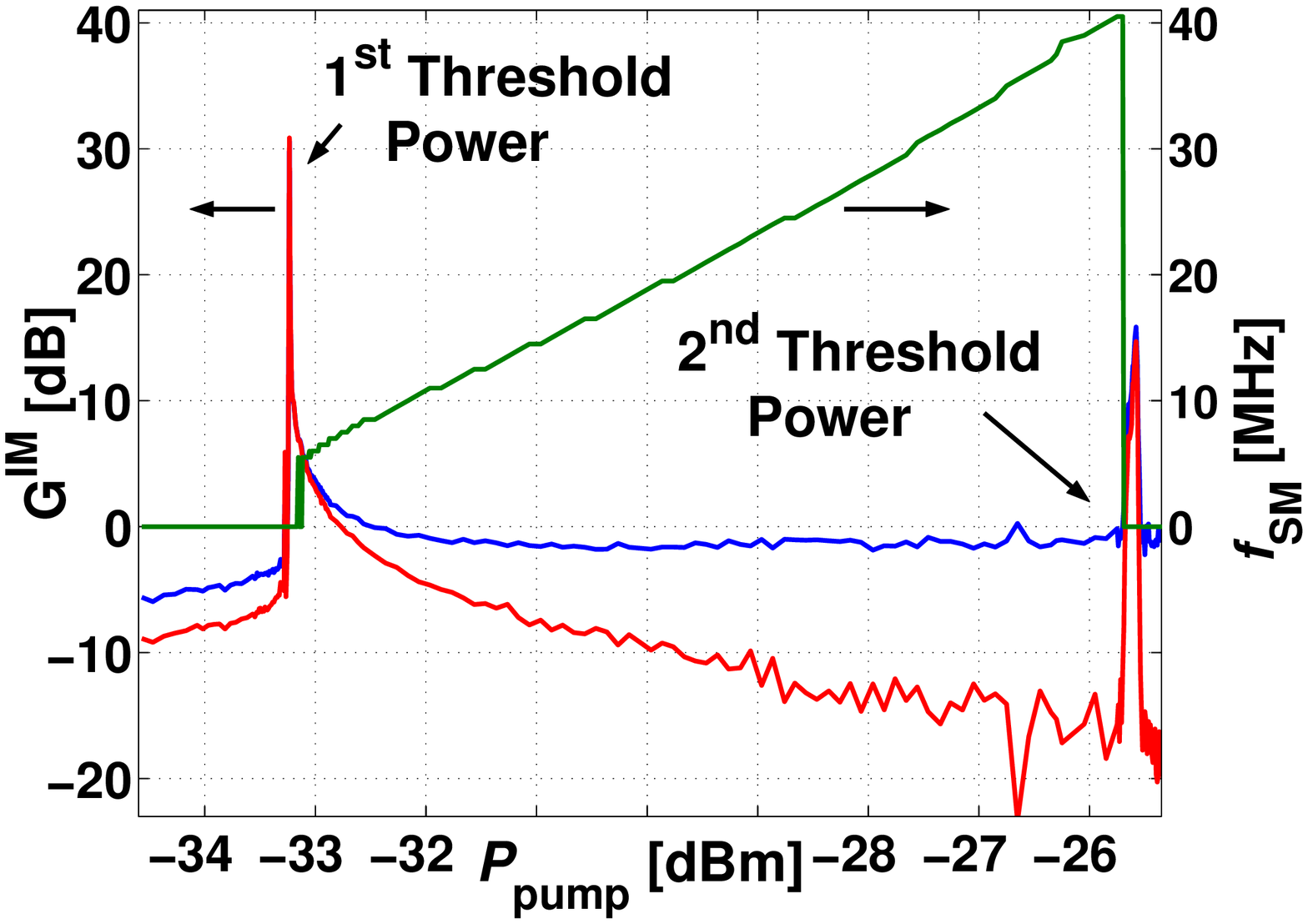}%
\\
{\protect\small (F)}%
\end{center}}}%
}%
\end{array}
\]
%

\caption{A typical IM measurement. $(\mathrm{A}%
)$ The reflected power $P_{\mathrm{refl}}$, is plotted as
a function of the measured frequency $f_{\mathrm{SA}}$,
centralized on the third resonance frequency $f_{\mathrm{3}}$
$(f_{\mathrm{SA}}^{\mathrm{c}}=f_{\mathrm{SA}}-f_{\mathrm{3}})$.
$(\mathrm{B}),(\mathrm{C}),(\mathrm{D})$ Pump,
Signal and Idler IM gains respectively, as a function of the input Pump and input Signal powers.
$(\mathrm{E})$ Simultaneous SM Measurement. The reflected power is plotted as
a function of the measured frequency centralized on $f_{\mathrm{3}%
}$ and the input Pump power, while the
resonator is stimulated with a single monochromatic tone at $f_{\mathrm{3}}$.
$(\mathrm{F}%
)$ Signal (blue) and Idler (red) IM gains versus the SM frequency (green) as a function
of input Pump power. }%
\label{IM_Graphs}%
\end{figure*}%

One of the common and effective ways to characterize nonlinearities in
superconductors is using IM measurements \cite{Baleegh06b}. IM, as measured in
our devices, is the result of two unequal tones, called Pump and Signal,
having close spaced frequencies, being mixed together by a nonlinear system,
which produces additional tones at frequencies that are linear combinations
(integer multiples) of both. The largest IM products appear at the third order
IM mixing tones, known as the Signal and Idler tones, because usually, the
second order mixing products do not coincide with any natural resonance
frequency of the resonator\cite{Nonlinear_Chin92}. The Idler is measured at
frequency $f_{\mathrm{Idler}}=2f_{\mathrm{pump}}-f_{\mathrm{sig}%
}=f_{\mathrm{pump}}-\Delta f$. Theory predicts that as the input pump tone
drives the resonator to the edge of instability, both the Signal and the Idler
tones undergo equally large amplification \cite{IM_Monaco2000}.

Fig. \ref{IM_Graphs}$(\mathrm{A})$ shows a typical IM measurement obtained at
the third resonance mode, namely $f_{\mathrm{pump}}=f_{\mathrm{3}}=5.667%
\operatorname{GHz}%
$. The Signal tone is deviated by $\Delta f=800%
\operatorname{Hz}%
$ from the Pump tone, and thus the Idler and both these tones lie within the
resonance band. The pump power is set to the edge of instability, namely to
the first SM power threshold at which the SM starts. In this region the Pump,
Signal and Idler tones as well as higher order mixing products are easily
detected in the reflected power spectrum. The higher order mixing products
also spread beyond the scope of this graph, and thus indicate the strength of
the IM nonlinearity at that region.\ 

The strength of the IM nonlinearity can be qualitatively characterized by
defining two parameters, IM Signal and Idler gains, defined as the ratio of
the output Signal and the output Idler powers to the input Signal power
respectively. Fig. \ref{IM_Graphs}$(\mathrm{C})$ and $(\mathrm{D})$ plot these
gains respectively as a function of both input Pump and input Signal powers.
Two areas of strong amplification, indicated by red shaded colors, are easily
noticed and referred to as first and second power thresholds. The maximum
amplifications achieved by the Signal tone at first and second threshold
powers are $30.66~$dB and $17.77~$dB, respectively. Whereas the maximum
amplifications achieved by the Idler tone are $30.85~$dB and $16.36~$dB,
respectively. As expected, the Signal and Idler gains are approximately equal.
To emphasize the strength of these amplifications we note that usually, no
amplification greater than unity ($0~$dB) is achieved in IM measurements with
superconducting resonators \cite{IM_Monaco2000,Nonlinear_Chin92}. Fig.
\ref{IM_Graphs}$(\mathrm{B})$ shows the corresponding gain of the Pump tone.
It experiences a strong absorption (pump depletion) at the first power
threshold, where the large amplification of the Pump and Signal tones takes
place. An increased absorption also occurs at the second power threshold, but
is harder to notice, as it is rather small on this scale.

In order to check the correlation between the IM and the SM phenomena, SM
measurement has been obtained simultaneously with the IM measurement, as shown
in Fig. \ref{IM_Graphs}$(\mathrm{E})$. At low input powers, approximately
below $-33.25~$dBm, and at high input powers, approximately above
$-24.5\ $dBm, the response of the resonator is linear, namely, the reflected
power off the resonator contains a single spectral component at the frequency
of the stimulating pump tone. In between the linear regions, there exists a
rather large power range, in which regular SM of the reflected power off the
resonator occurs. It is realized by rather strong and sharp sidebands, which
can extend for several hundred megahertz to both sides of the resonance
frequency. The SM frequency, defined as the frequency difference between the
pump and the primary sideband, increases as the pump power increases. The
regular SM starts and ends at two power thresholds. The first threshold occurs
on a very narrow power range of approximately $10%
\operatorname{nW}%
$, where the resonator response desists from being linear. It experiences a
strong amplification of the noise floor (noise rise), over a rather large
frequency band, especially around the resonance frequency itself. The second
power threshold occurs on a slightly larger power range than the first one and
has similar, but less extreme characteristics. As expected, the noise rise
during the regular SM is negligible, as the transition through the instability
point is fast \cite{segev06c,BifAmp_Kravtsov01}.

Fig. \ref{IM_Graphs}$(\mathrm{F})$ shows the correlation between the SM and
the IM phenomena. The blue (red) curve shows the IM Signal (Idler) gain as a
function of the input pump power, while the input signal power equals
$-86~$dBm ($-79.6~$dBm), for which the maximum of the gain is measured. The
green curve shows the SM frequency as a function of the input pump power. A
comparison clearly shows that the strong IM gain is achieved at the same power
thresholds of the SM phenomenon. This finding indicates that the correlation
between these two phenomena is strong, and proves that the observed IM does
not originate from various other possible nonlinear mechanisms, common in
superconductors\cite{Baleegh06a}. The narrow power range, at which the Signal
and Idler gains are achieved, which is by most $5%
\operatorname{nW}%
$ wide, emphasizes the sharpness of the edge of instability in our device.
\begin{figure*}
\centering
\[
\text{%
{\includegraphics[
height=3.521in,
width=4.5778in
]%
{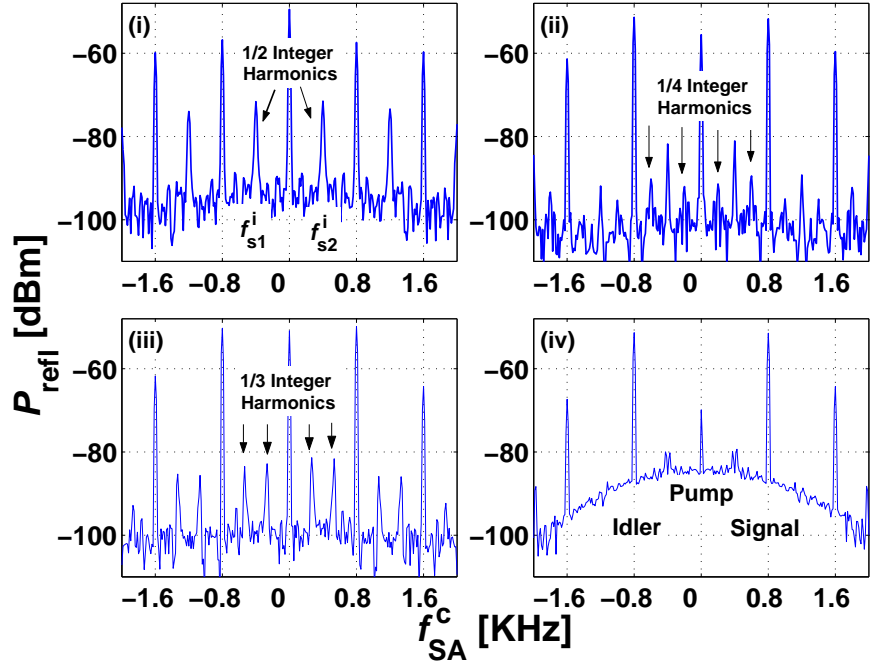}%
}%
}%
\]
%

\caption{Period doubling captured during
IM measurement. The reflected power is plotted as a function of the measured
frequency $f_{\mathrm{SA}}$ centralized on $f_{3}$ $($ $f_{\mathrm{SA}%
}^{\mathrm{c}}=f_{\mathrm{SA}}-f_{3})$. It shows the Pump, Signal, and Idler
tones, and in addition mixing products of $(i)$ half, $(ii)$ quarter, and $(iii)$ one third subharmonics of the Pump and Signal tones. Panel $(iv)$
shows chaotic-like behavior, characterized by a strong and broadband
amplification of the noise floor.}%
\label{e13-PeriodDoubling}%
\end{figure*}%

It has been found that in some nonlinear systems, the transition to chaos can
occur via consecutive PDB instabilities of various orders
\cite{perDoub_May76,perDoub_Feigenbaum78,perDoub_Linsay81,perDoub_Balberg94}.
It was shown that near the onset of PDB, any dynamical system can be used to
amplify perturbations near half the fundamental frequency. The closer the
bifurcation point is, the greater is the amplification
\cite{perDoub_Wiesenfeld85b}. During IM measurement, PDB is indirectly
observed at some of our devices, at the SM power threshold. Naturally, the
sub-harmonics that result from PDB do not coincide with other modes of the
resonator and therefore can not be directly measured. The observation of PDB
is possible because the sub-harmonics of the Pump and Signal tones mix
together and appear at new frequencies in the vicinity of the stimulated
resonance mode.

The amplification caused by the PDB, which occurs at rational fractions of the
resonance frequency, has a great advantage over amplifications of tones which
lie in the resonance band. The latter tones are always perturbed by the
resonance mode, whereas the former tones can be effectively decoupled from the
resonance mode by setting the input pump off the power threshold. The coupling
occurs only during the time periods at which a measurement is deliberately
obtained. Thus our system has both the ability of strong amplification and can
be coupled or decoupled at will, to a measured system, having signals at half
the resonance frequency of the resonator.

Fig. \ref{e13-PeriodDoubling} has four subplots, each showing an IM
measurement, in which PDB\ is observed. The strong Pump and weak Signal powers
are set to the SM power threshold, and the reflected Pump, Signal, Idler, and
high order mixing tones, are easily observed (labeled at subplot $(iv)$). In
addition, looking at subplot $(i)$, we observe a new type of reflected tones
which are found at half integer multiples of the Pump and Signal frequencies.
For example, the two labeled tones are found at $\left(  f_{\mathrm{pump}%
}+f_{\mathrm{sig}}\right)  /2$,$~$and $\left(  3f_{\mathrm{pump}%
}-f_{\mathrm{sig}}\right)  /2$. This measurement provides a clear evidence
that a period doubling of the second order occurs in the resonator. Subplots
$(ii)$ and $(iii)$ show additional measurements, in which period doubling of
forth and third orders occur, respectively. Subplot $(iv)$ shows a measurement
in which a chaotic-like behavior is observed. It is characterized by a strong
and broadband amplification of the noise floor and a high absorption of the
pump power. At first these measurements were taken while sweeping the weak
Signal power, but later we have realized that due to a quasi-periodic, very
low frequency noise the system can self-switch between the various period
doubling states even when all deliberately applied external excitations are
constant. At this point the origin of this low frequency noise is yet
unclear.
\begin{figure*}
\centering
\[%
\begin{array}
[c]{cc}%
\text{%
{\parbox[b]{2.1967in}{\begin{center}
\includegraphics[
height=1.1523in,
width=2.1967in
]%
{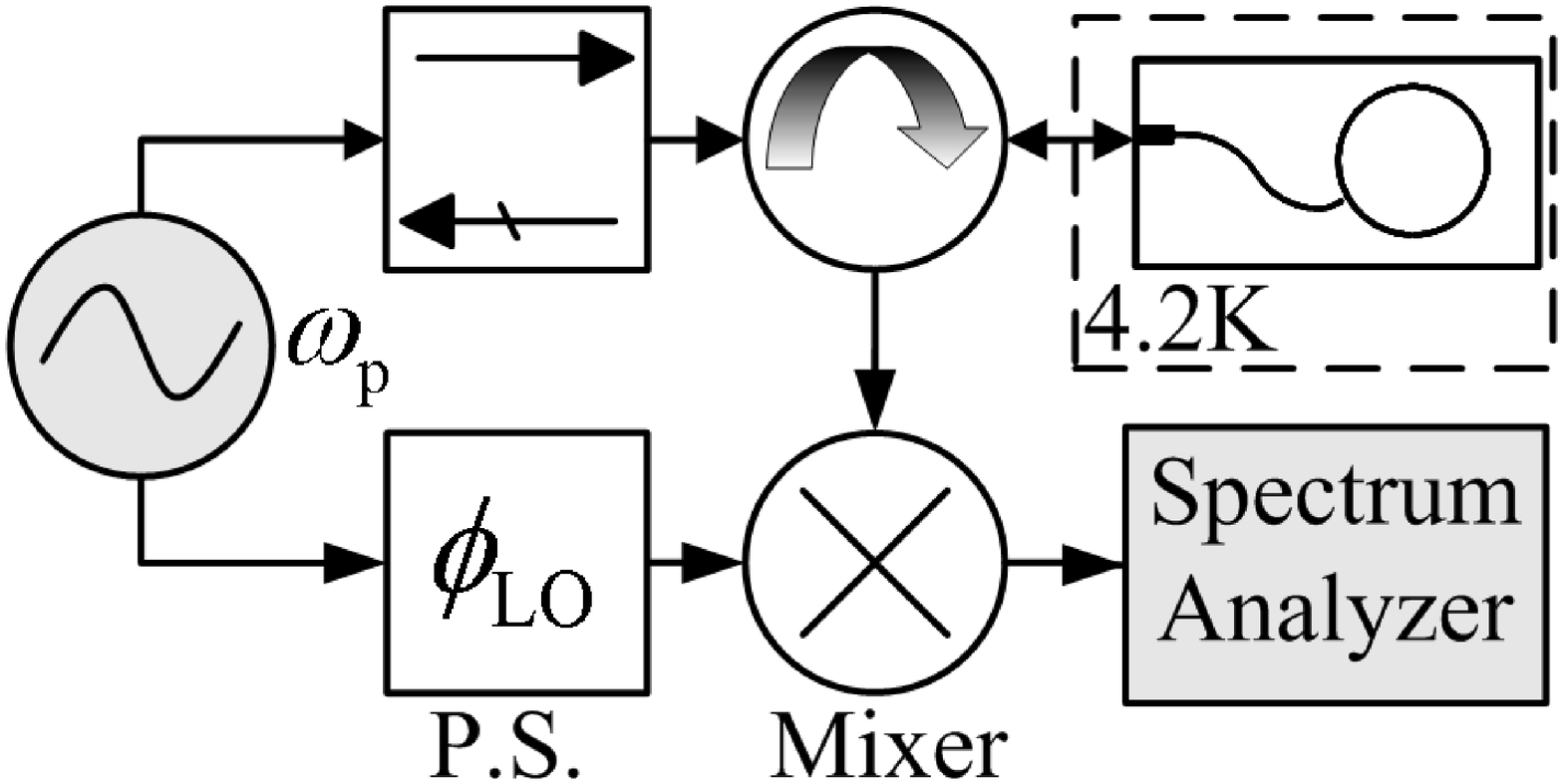}%
\\
{\protect\small {}(A)}%
\end{center}}}%
} & \text{%
{\parbox[b]{2.1959in}{\begin{center}
\includegraphics[
height=1.7069in,
width=2.1959in
]%
{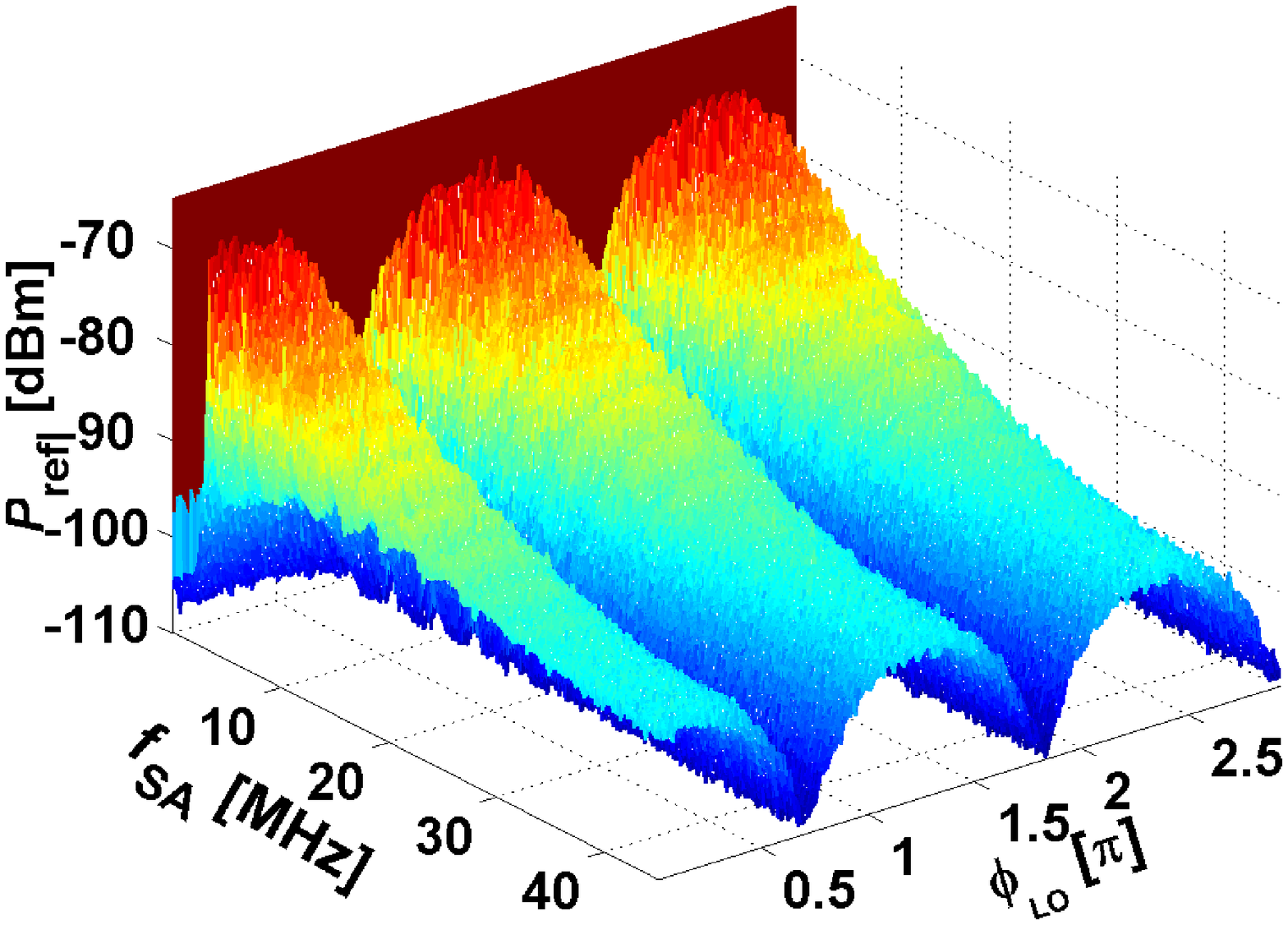}%
\\
{\protect\small (C)}%
\end{center}}}%
}\\
\text{%
{\parbox[b]{2.1685in}{\begin{center}
\includegraphics[
height=1.1067in,
width=2.1685in
]%
{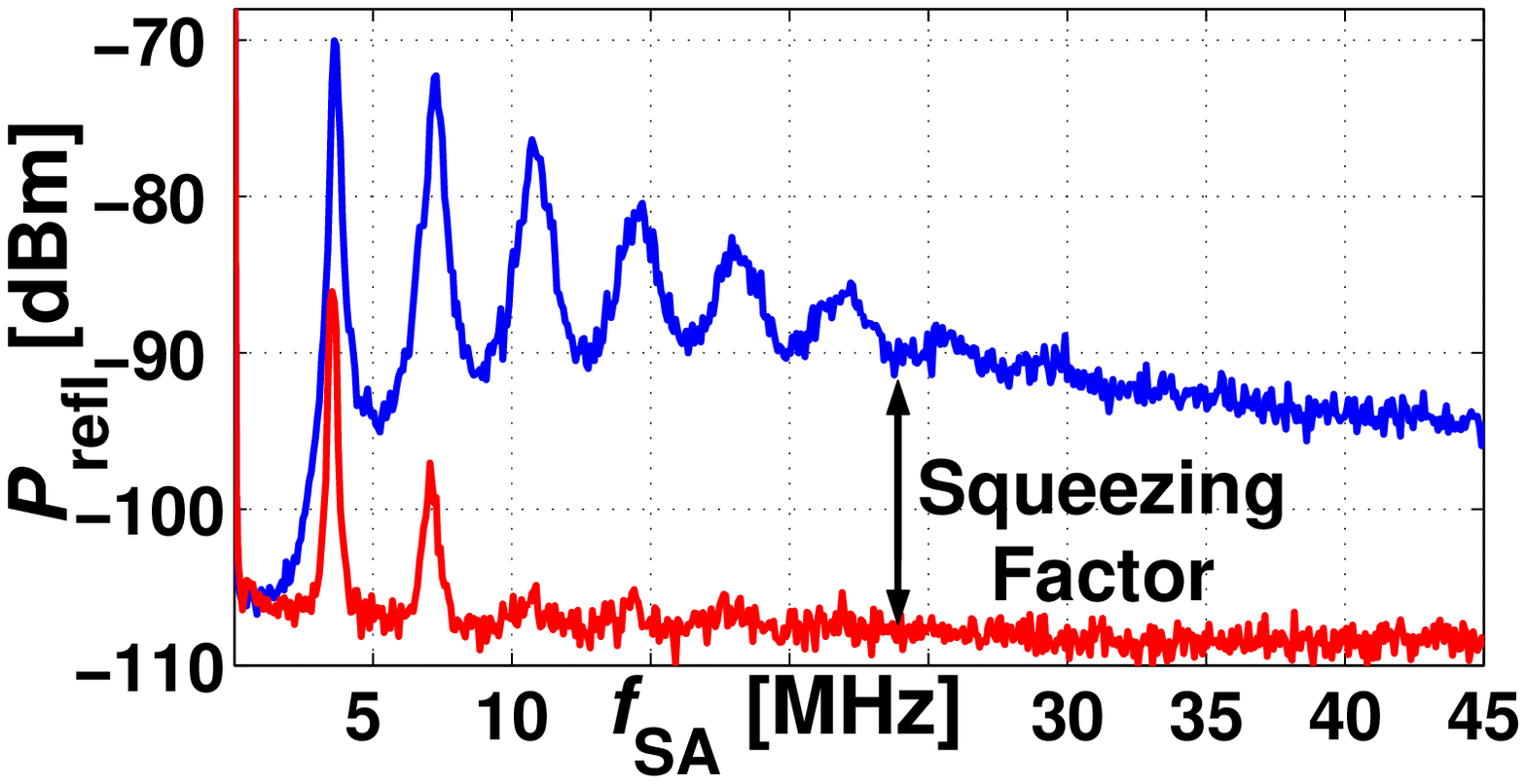}%
\\
{\protect\small (B)}%
\end{center}}}%
} & \text{%
{\parbox[b]{2.1959in}{\begin{center}
\includegraphics[
height=1.6887in,
width=2.1959in
]%
{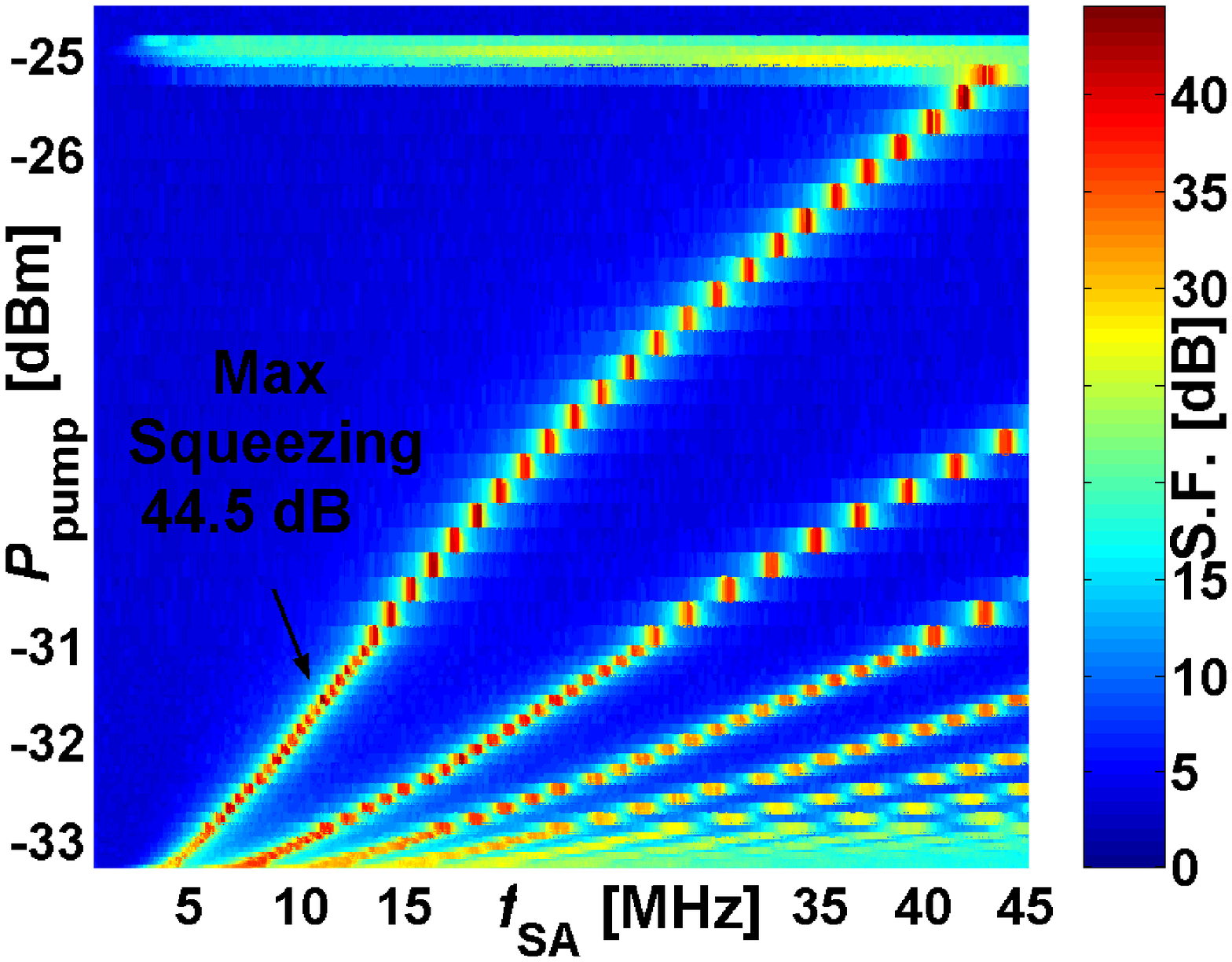}%
\\
{\protect\small (D)}%
\end{center}}}%
}%
\end{array}
\]
%

\caption{$(\mathrm{A}%
)$ PSD measurement setup, which utilizes homodyne detection measurement scheme.
The resonator is stimulated by a single monochromatic pump and the reflected power is fed to an external mixer,
driven by a local oscillator (LO), having the same frequency as the pump, and an adjustable phase.
The output of the mixer is measured by a SA in a frequency band of $45 $MHz starting from DC.
$(\mathrm{B})$ A typical PSD measurement. Reflected power as a function of
the measured frequency for $\phi_{\mathrm{LO}}=0$ (blue) and  $\phi
_{\mathrm{LO}}=\protect\pi/2$ (red).
The same measurement for a continuous LO phase is shown in $(\mathrm{C})$.
$(\mathrm{D}%
)$ Squeezing factor as a function of the measured frequency and the input pump power.}%
\label{SQZ_Graph}%
\end{figure*}%

A parametric amplifier can establish correlations \cite{Sqz_Yurke85} between
the output tones at $f_{\mathrm{pump}}\pm\Delta f$, where $\Delta f=\left\vert
f_{\mathrm{sig}}-f_{\mathrm{pump}}\right\vert $, as a result of IM gain. When
delivered to a mixer, operated in the homodyne mode, whose local oscillator
(LO) is phase-locked to the pump, these correlations can result in noise
fluctuations reduced below that which the mixer would see if the signal
delivered to the parametric amplifier were, instead, directly delivered to the
mixer \cite{Squeezing_Yurke05}. Consequently, the measured spectral power
becomes $\pi$-periodical, as a function of the phase $\phi_{\mathrm{LO}}$ of
the LO, relative to the pump. This phenomenon is called PSD \cite{sqz_Bocko88}
or squeezing for thermal noise\cite{Sqz_Yurke88}, weak
signals\cite{Sqz_Almog06} or quantum fluctuations \cite{Sqz_Movshovich90}
reduction. The theory is detailed in \cite{Sqz_Yurke85}, and summarized in
\cite{Sqz_Yurke89}. In our resonators we observe PSD of both the fluctuation
noise and the spectral power density of the sidebands generated by the SM phenomenon.

Typical PSD measurement, as obtained using the setup described in Fig.
\ref{SQZ_Graph}$(\mathrm{A})$, is shown in Fig. \ref{SQZ_Graph}$(\mathrm{B})$.
The blue curve shows the largest amplification of the measured spectrum. The
pump tone is down converted to dc and the SM sidebands are down-converted
respectively to the pump. We refer to this measurement as taken with a zero LO
phase, $\phi_{\mathrm{LO}}=0$. The red curve is taken with $\phi_{\mathrm{LO}%
}=\pi/2$. The deamplification of the noise, relatively to $\phi_{\mathrm{LO}%
}=0$ curve is clear. Panel $(\mathrm{C})$ shows a similar measurement, taken
for continuous $\phi_{\mathrm{LO}}$ values. The dependence of the reflected
power on $\phi_{\mathrm{LO}}$ is clearly observed, where the phase period
equals $\pi$, as expected. We define the squeezing factor as the ratio of the
measured signal at its maximum value to the measured signal at its minimum
value. Panel $(\mathrm{D})$ shows the squeezing factor as a function of the
measured frequency and the pump power. Strong and broadband squeezing occurs
at the two power thresholds, where the strong IM gain is measured. In
addition, an even stronger squeezing occurs at the SM sidebands where a
maximum value of $44.5~$dB has been measured.

In conclusion, our devices exhibit an extreme nonlinear behavior which
manifests itself in various effects, in correlation with the SM phenomenon.
Strong IM gain of both the Signal and Idler tones is measured in all of our
devices at the SM power threshold. Correlated PDB of various orders as well as
chaotic-like behavior occur at some of our devices. In addition, PSD is
measured, and shows a strong squeezing factor in correlation with the IM
strong gain. The outcome of this three correlated phenomena is a very strong,
phase sensitive amplifier, which can be coupled or decoupled to a measured
system, at will. Such a device is highly suitable for quantum data processing
systems and for quantum state readout schemes.

\begin{acknowledgement}
We thank Bernard Yurke, Ron Lifshitz, Mile Cross, Oded Gottlieb, and Steven
Shaw for valuable discussions. This work was supported by the German Israel
Foundation under grant 1-2038.1114.07, the Israel Science Foundation under
grant 1380021, the Deborah Foundation, the Poznanski Foundation, and MAFAT.
\end{acknowledgement}

\bibliographystyle{Science}
\bibliography{Bibilography}

\end{document}